\documentclass[12pt,english]{iopart}
\usepackage{iopams}
\usepackage{mathptmx}
\usepackage{graphicx}
\usepackage{amssymb}
\usepackage{babel}
\usepackage[usenames,dvipsnames]{color}
\usepackage[seperr, load={}]{siunitx} 

\newcommand*{\Pdet}{P_{\rm det}}

\newcommand*{\stos}{|s_{21}|^2}
\newcommand*{\Tb}{T_{\rm bath}}
\newcommand*{\Te}{T_{\rm e}}
\newcommand*{\kB}{k_{\rm B}}

\newcommand*{\Vb}{V_{\rm b}}
\def\be{\begin{equation}}
\def\ee{\end{equation}}

\newcommand*{\RT}{R_\mathrm{T}}

\newcommand*{\Gth}{G_{\rm th}}
\newcommand*{\Pin}{P_{\rm in}}
\newcommand*{\Pout}{P_{\rm out}}

\newcommand*{\Tsys}{T_{\rm sys}}
\newcommand*{\Cin}{C_{\rm C1}}
\newcommand*{\Cout}{C_{\rm C2}}

\usepackage{color}
\definecolor{dred}{rgb}{.6,.0,0.}
\definecolor{dblue}{rgb}{.0,.0,0.6}

\begin{document}

\title[Incomplete measurement and work]{Incomplete measurement of work
in a dissipative two level system}

\author{Klaara Viisanen$^1$, Samu Suomela$^2$, Simone Gasparinetti$^{1}$, Olli-Pentti Saira$^1$, Joachim Ankerhold$^{3, 4}$, and Jukka P. Pekola$^1$}
\address{$^1$ Low Temperature Laboratory, Department of Applied Physics, Aalto University School of Science, POB 13500, FI-00076 Aalto, Finland}
\address{$^2$ Department of Applied Physics and COMP Center of Excellence, Aalto University School of Science, P.O. Box 11100, FI-00076 Aalto, Finland}
\address{$^3$ Institute for Complex Quantum Systems, Ulm University, 89069 Ulm, Germany}
\address{$^4$ Center for Integrated Quantum Science and Technology (IQST), 89069 Ulm, Germany}

\date{\today}

\begin{abstract}
We discuss work performed on a quantum two-level system coupled to multiple thermal baths. To evaluate the work, a measurement of photon exchange between the system and the baths is envisioned. In a realistic scenario, some photons remain unrecorded as they are exchanged with baths that are not accessible to the measurement, and thus only partial information on work and heat is available. The incompleteness of the measurement leads to substantial deviations from standard fluctuation relations. We propose a recovery of these relations, based on including the mutual information given by the  counting efficiency of the partial measurement. We further present the experimental status of a possible implementation of the proposed scheme, i.e. a calorimetric measurement of work, currently with nearly single-photon sensitivity. 

\end{abstract}
\setcounter{tocdepth}{2}
\tableofcontents



\maketitle

\section{Introduction}\label{intro}
The study of nonequilibrium thermodynamics in quantum systems has witnessed fast progress in the last decade. Especially,  theoretical advancements have been achieved not only for closed systems but also for open quantum systems \cite{esposito09,campisi11}. However to this day, the measurement of thermodynamic quantities, such as work, in coherent quantum systems  has been limited to unitary dynamics in the experiments \cite{paternostro14}. Although several techniques have been proposed \cite{huber2008,heyl2012,pekola13,dorner2013,mazzola2013,campisi13}, the interesting case presented by open quantum systems is still to be explored experimentally. For such an experiment to be possible, one needs to monitor all the relevant degrees of freedom, including the environment. This approach would reduce the dynamics again to that of a closed system comprised of the quantum system itself together with its environment. One of the possible schemes in this direction is a calorimetric measurement of the relevant environment \cite{pekola13}. In such a measurement, energy is detected as temperature variation in an absorber with low heat capacity. Ideally, for a two-level system such a measurement yields all the relevant information, including the initial and final states of the system itself.

The topic of this article is to assess quantitatively how the {\sl counting efficiency} of such a measurement influences its outcome in terms of work and its distribution. We define the counting efficiency as the number of photons detected divided by the total number of photons exchanged. In the case of a "hidden", unmeasured environment at the same temperature as the measured one, the results become particularly simple. Analytical results can be obtained in the standard situation where the system is coupled to the reservoirs only before and after the driving period. We recover the fluctuation relations once we include in them mutual information, which directly relates to the counting efficiency of the measurement.  The quantum trajectory (quantum jumps) method yields numerical answers in the general case of a qubit coupled to the reservoirs also during the application of the driving protocol. 

We additionally provide an update on the progress made in the implementation of the calorimetric measurement toward a single-microwave-photon detection. The first steps in implementing the calorimetric measurement experimentally have been reported elsewhere \cite{nahum94,nahum95,schmidt03,schmidt04,schmidt05,gasparinetti14}. Continuing the work started in Ref. \cite{gasparinetti14}, we report significantly improved results in terms of the measurement noise. This method presents a promising way for the proposed studies in the near future. In such a measurement the counting efficiency would be determined mainly by the intrinsic decay of the qubit to the "dark" environments, determined by the relaxation time of it in the absence of the engineered calorimeter. 

\section{Preliminaries}\label{preliminaries}

We consider a quantum system with Hamiltonian $H_S(t)=H_0+H_D(t)$, where $H_D(t)$ describes an external time-dependent drive in the interval $t\in [t_i, t_f]$ with $H_D(t_i)=H_D(t_f)=0$. This system is embedded in a dissipative reservoir described by $H_R$ so that the total compound is captured by
\begin{equation}\label{hamiltonian}
H(t)=H_S(t)+H_I+H_R,
\end{equation}
with $H_I$ being the  interaction part. While driven open systems have been studied extensively in the past, our focus here lies on the measurement of the work exerted by the drive on the system in presence of dissipation. Since work itself is not a proper quantum observable, the calculation of its distribution must be performed with care \cite{campisi11}.

\subsection{Work and dissipative dynamics}

A consistent formulation of work in a closed system is provided by the two measurement protocol (TMP) \cite{kurchan00,talkner07} which even allows to retrieve the full distribution of work \cite{esposito09,campisi11}. According to this scheme, the probability to measure energy $E_i$ at time $t=t_i$ and $E_f$ at time $t=t_f$ and thus the probability distribution for the work $W=E_f-E_i$  is given by
\begin{equation}\label{distribution}
p(W) \equiv p(E_f-E_i) ={\rm Tr}\{ \Pi_f\, U(t_f,t_i)\, \Pi_i \mathcal{W}(t_i)\, \Pi_i \, U^\dagger(t_f,t_i)\, \Pi_f\}\, ,
\end{equation}
where $U(t_f,t_i)=\mathcal{T} \exp[-\frac{i}{\hbar} \int_{t_i}^{t_f} dt H(t)]$ is the unitary time evolution operator, $\Pi_{i/f}=|E_{i/f}\rangle \langle E_{i/f}|$ are projection operators on energy eigenstates at the initial and final time, respectively, and $\mathcal{W}(t_i)$ is the initial equilibrium density with respect to $H(t_i)=H_0+H_I+H_R$. The $k-$th moment of work easily follows as
\begin{equation}
\langle W^k\rangle = \int dw \, w^k \delta[w-(E_f-E_i)]\, P[E_f, E_i]\, .
\end{equation}

However, for dissipative  systems this formulation is difficult if not impossible to implement in an actual experiment due to the fact that the reservoir degrees of freedom are neither accessible nor controllable. To perform projective measurements on eigenstates of the full compound is thus not feasible. As long as one is interested only in the first and second moment of work, one may alternatively consider the power operator \cite{solinas13}
\begin{equation}\label{poweroperator}
  {P}_W(t)=\frac{\partial H_S(t)}{\partial t}\, .
\end{equation}
The time integrated moments of its corresponding Heisenberg operator provide results identical to those obtained from (\ref{distribution}) if expectation values are taken with respect to thermal initial states \cite{suomela14}. In the regime of weak system-reservoir interaction and sufficiently weak driving, these moments can be obtained based on the time evolution of the reduced density $\rho(t)={\rm Tr}_R\{ \mathcal{W}(t)\}$, i.e.,
\begin{equation}
\dot{\rho}(t)=-\frac{i}{\hbar} [H_0+H_D(t),\rho(t)]+\mathcal{L}[\rho],
\end{equation}
with the dissipator $\mathcal{L}$ determined by reservoir induced excitation and emission rates $\Gamma^{\downarrow , \uparrow}$ related to each other by detailed balance. A simple calculation using the power operator (\ref{poweroperator}) then leads to the first law of thermodynamics $\langle W\rangle = \langle \Delta U\rangle + \langle Q \rangle $, with the work being the sum of the change in internal energy and the heat flow. Here and in the following we use the sign convention  
that for heat flow into (out of) the reservoir $Q>0$ ($Q<0$).

\subsection{Probing the reservoir}

To make progress on more general grounds, it has been proposed to evaluate work by monitoring directly the energy exchange between system and reservoir \cite{campisi11,pekola13}. In the regime of weak coupling between a system and its surrounding this then provides the work statistics performed on the open system. Theoretically, this scheme is conveniently implemented within the so-called quantum jump (QJ) formulation. An alternative route is provided by generalized master equations \cite{esposito09,silaev14,gasparinetti14NJP}. The QJ method has been pioneered in quantum optics to describe emission and absorption processes of single photons by few level systems (atoms) \cite{dalibard92}. The method exploits the probabilistic nature of the quantum mechanical time evolution by constructing the dynamics $|\psi(t)\rangle \to |\psi(t+\Delta t)\rangle$ over a time interval $\Delta t$ according to sequences of jumps between energy levels with transition probabilities determined by the corresponding Hamiltonian \cite{plenio98,carmichael93}. Practically, one uses a Monte Carlo procedure to sample individual quantum trajectories, and the distribution is obtained by averaging over a sufficiently large number of realizations.

This method has recently been formulated to record the exchange of energy quanta between a two level system (TLS)
\begin{equation}\label{tls}
H_S(t)=\frac{\hbar\omega_0}{2}\, \sigma_z +\lambda(t)\, \sigma_x,
\end{equation}
with $\sigma_x, \sigma_z$ being Pauli matrices, $\hbar\omega_0$ the level spacing, and $\lambda(t)$ the external driving field \cite{hekking13}. The idea is to count the last photon before the drive starts and the first phonon exchanged after the drive ends. These "guardian photons" can be used to detect the respective states of the TLS and thus to retrieve information about the change in internal energy. On the other hand, monitoring the photon exchange during the drive provides the net heat flow. As long as the weak coupling assumption applies, the sum of these two quantities provides the work. Experimentally, this information is obtained by a calorimetric measurement of the heat bath if an energy resolution on a single photon level is achieved.

\section{Incomplete measurement for a driven two level system}\label{incomplete}

In order for a heat bath to function as an efficient detection medium, its energy exchange with the system must be fully under control. Typically, however, only parts of the environment interacting with a system of interest are known and calorimetrically accessible. Other components remain unidentified while still influencing the system. Assuming that all components can be considered as independent heat baths, one can extend the model (\ref{hamiltonian}) by putting $H_R=H_{R, probe}+H_{R, dark}$ with $H_{R, probe}$ being the part which can be probed and $H_{R, dark}$ accounting for the unobserved heat baths. Monitoring $H_{R, probe}$ thus delivers only partial information about the state of the system before, during, and after the drive. The question is then to what extent a corresponding measurement provides information about the work statistics. 

The QJ approach can treat this problem numerically. Analytical insight is obtained by neglecting the photon exchange during the drive (very weak system-baths coupling) and focusing on the counting efficiency of detecting the correct result for the initial and final states of the TLS.
\begin{figure}[t]
\centerline{\includegraphics[width=.6\columnwidth]{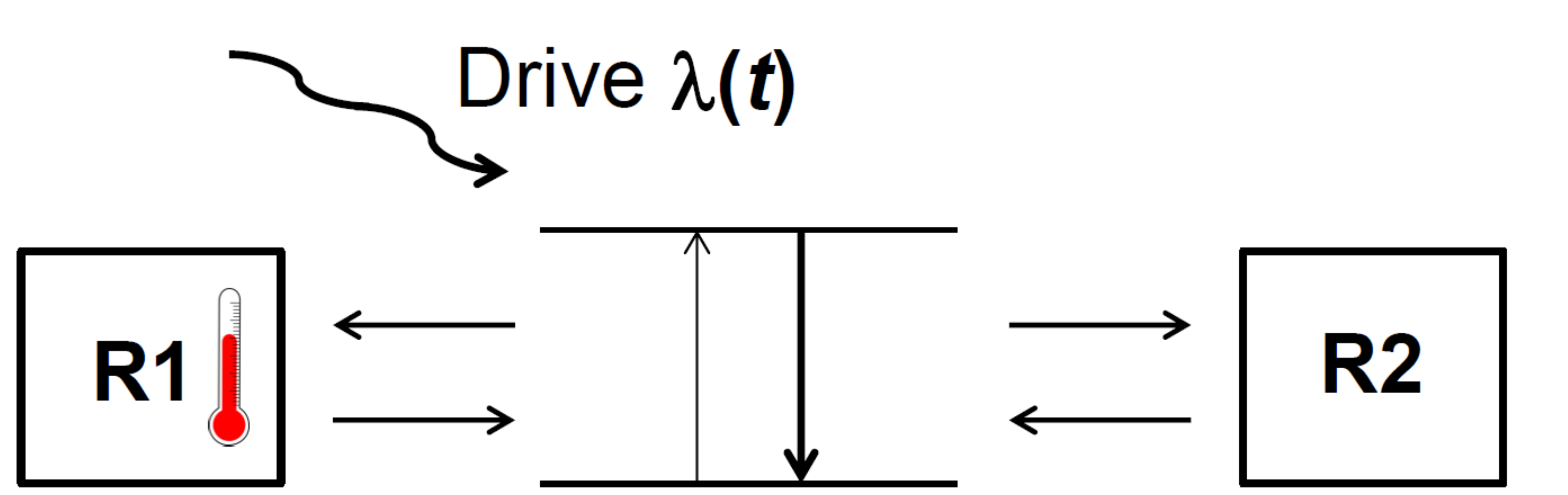}}
\caption{TLS (centre) coupled to two reservoirs (R1 and R2), subject to a time dependent drive $\lambda(t)$. Only the reservoir R1 is calorimetrically measured.}
\label{figure1}
\end{figure}

\subsection{Incomplete work measurement}
We consider a setup where a TLS is embedded into two independent heat baths $H_{R, probe}$ and $H_{R, dark}$, where only the first one is measured calorimetrically. The setup is schematically illustrated in Fig. \ref{figure1}. Both environments are assumed to be at the same temperature $k_{\rm B} T=1/\beta$ and to interact only very weakly with the TLS. The relevant quantity is then the relative strength of the couplings between TLS and $H_{R, probe}$ and $H_{R, dark}$, i.e.,
\begin{equation}
\eta=\frac{\Gamma^\downarrow_{probe}}{\Gamma^\downarrow_{probe}+\Gamma^\downarrow_{dark}}=
\frac{\Gamma^\uparrow_{probe}}{\Gamma^\uparrow_{probe}+\Gamma^\uparrow_{dark}},
\end{equation}
with emission/excitation rates corresponding to the respective reservoirs. The counting efficiency $\eta$ gives the probability of photon emission/absorption between the TLS  and the probe reservoir, while $1-\eta$ is the probability that the quantum is exchanged with the dark reservoir. For example, $(1-\eta)\eta$ is the probability of predicting erroneously the TLS to be in the excited state after the absorption of one photon from the probe reservoir, while its true state is the ground state due to a subsequent emission of a photon into the dark reservoir. The probability $\epsilon$ of making an error in determining the initial (final) state before (after) the drive is obtained by summing up all unobserved higher order events
\begin{equation}
\epsilon =(1-\eta)\, \eta \, \sum_{k\geq 0} (1-\eta)^{2k}= \frac{1-\eta}{2-\eta}\, .
\end{equation}
Likewise, the probability to predict the state of the TLS correctly by measuring the probe reservoir is given by
\begin{equation}
1-\epsilon = \eta \sum_{k\geq 0} (1-\eta)^{2k}=\frac{1}{2-\eta}\, .
\end{equation}
Apparently, one regains an ideal detection for $\eta\to 1$, while the outcome predicts the true state of the TLS only with
probability $1-\epsilon=1/2$ for $\eta\to 0$.

Initially (before the drive) the TLS (\ref{tls}) is assumed to be in thermal equilibrium so that due to the weak coupling its probability to be in the ground state ('0') or in the excited state ('1') is given by
\begin{equation}
P(0)=1-P(1)=\frac{1}{1+{\rm e}^{-\beta\hbar\omega_0}}=\frac{1}{Z}\, {\rm e}^{\beta\hbar\omega_0/2}\, ,
\end{equation}
with partition function $Z=2 {\rm cosh}(\beta\hbar\omega_0/2)$. Prior to the drive, the actual state of the TLS is measured by the probe reservoir according to the above description. Starting from an energy eigenstate of the TLS the subsequent drive 
generates a unitary time evolution followed by a final measurement of the TLS via the probe reservoir. The expectation value of a function $f(W)$ of the work in this process is thus given by
\begin{equation}
\langle f(W)\rangle_\epsilon = \sum_{k_i, k_f=0,1} f[\hbar\omega_0 (k_f-k_i)]\, P_D(k_f, k_i),
\end{equation}
with 
\begin{eqnarray}\label{workdetect}
P_D(k_f, k_i)&=&\sum_{k, k'=0,1}[(1-\epsilon) \delta_{k', k_f}+\epsilon (1-\delta_{k', k_f})]\, p(k\to k')\nonumber\\
 &&\hspace{1cm}\times [(1-\epsilon) \delta_{k, k_i}+\epsilon (1-\delta_{k, k_i})]\, P(k),
\end{eqnarray}
where $P_D(k_f, k_i)$ is the detector probability to predict the TLS to be initially in state $k_i$ and to be finally in state $k_f$ if starting from the thermal distribution $P(k)$ and evolving during the drive with probability $p(k\to k')$ from state $k$ into state $k'$. For an ideal measurement $\epsilon=0$ this expression reduces to the expected result
\begin{equation}
\left. P_D(k_f,k_i)\right|_{\epsilon=0}= p(k_i\to k_f) P(k_i)\, .
\end{equation}

A basic example is the response of the TLS to a so-called $\pi$-pulse such that drive amplitude and duration swap the state, i.e.\ $p(k\to k')=1-\delta_{k k'}$. The above expressions then simplify to
\begin{eqnarray}\label{workaverage}
\langle f(W)\rangle_\epsilon &=& (1-\epsilon)^2 \left[ P(0) f(\hbar\omega_0) + P(1) f(-\hbar\omega_0)\right] + 2 (1-\epsilon)\epsilon f(0)\nonumber\\
&&+\epsilon^2 \left[ P(0) f(-\hbar\omega_0) + P(1) f(\hbar\omega_0)\right]\, .
\end{eqnarray}
\begin{figure}[b]
\centerline{\includegraphics[width=.6\columnwidth]{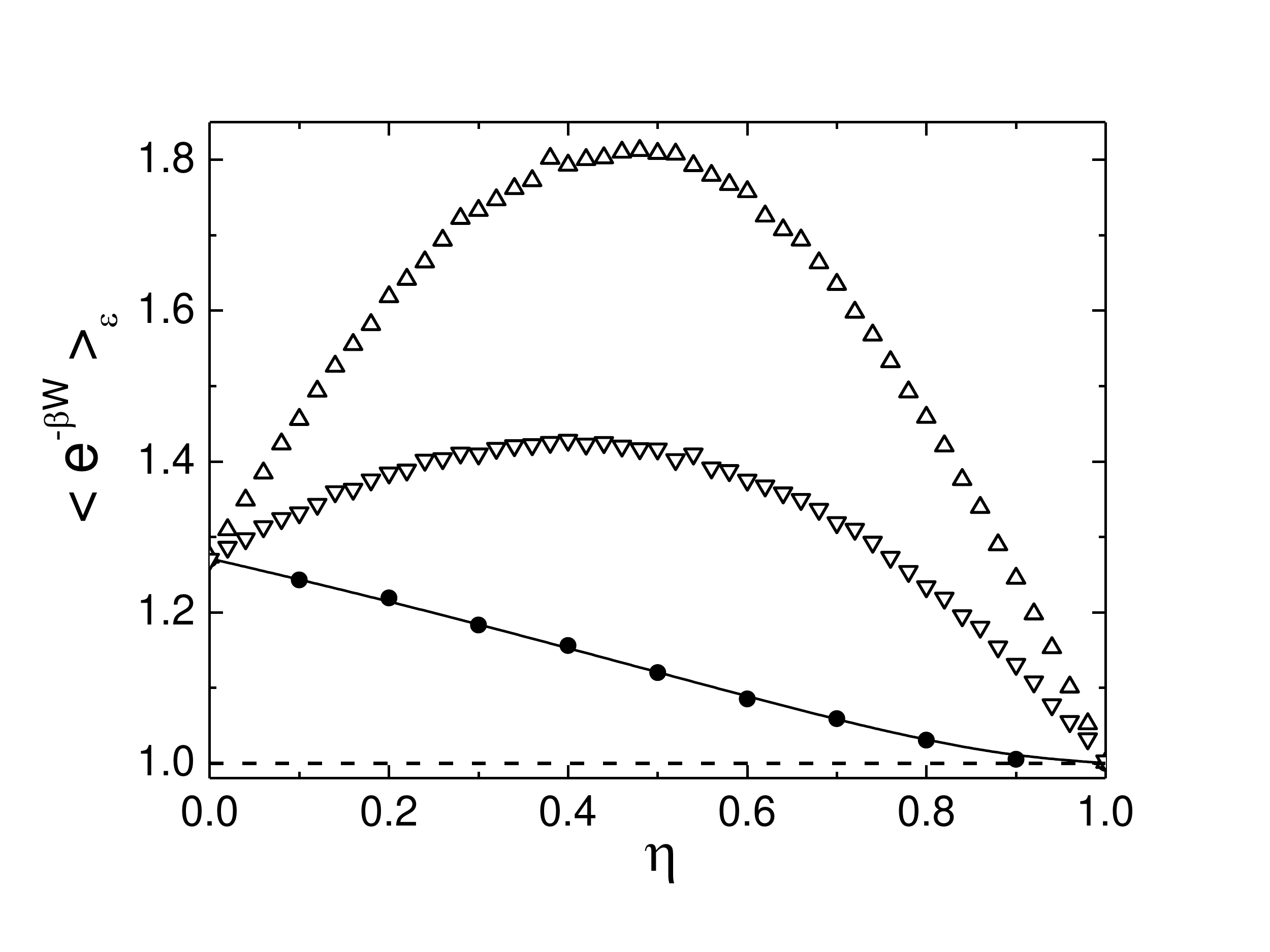}}
\caption{Deviations from Jarzynski relation for an incomplete measurement with two reservoirs having identical couplings to the TLS. The parameters are: $\beta \hbar\omega_0 = 1$, $\lambda(t) = \lambda_0 \sin(\omega_0 t)$, where $\lambda_0 = 0.05 \hbar\omega_0$, and the drive lasts over 10 periods ($\pi$-pulse). The different sets correspond to $\Gamma^\downarrow = 0$ (filled circles, numerical, and the solid line, analytic result of Eq. (\ref{je})), $\Gamma^\downarrow = 0.05\hbar\omega_0$ (down-triangles, numerical), and $\Gamma^\downarrow = 0.10\hbar\omega_0$ (up-triangles, numerical).}
\label{figure2}
\end{figure}

\subsection{Modified Jarzynski and Crooks relations}
By choosing $f(W)={\rm e}^{-\beta W}$ one arrives at a modified Jarzynski relation \cite{jarzynski97} of the form
\begin{equation} \label{je}
\langle {\rm e}^{-\beta W}\rangle_\epsilon=1+ 2 \epsilon^2\, {\rm e}^{\beta\hbar\omega_0/2}\,  {\rm sinh}(\beta\hbar\omega_0/2)\, .
\end{equation}
The ideal detection $\epsilon\to 0$ again provides the conventional result, while strong deviations occur for finite $\epsilon$ and especially at low temperatures. The deviation from the ideal result is always positive implying that the balance between work put into the system ($W>0$) and work extracted from the system ($W<0$) seems to be distorted in favor of these latter processes: This is due to wrong initial and final measurements, where the TLS is initially assumed to be in state 1 [while it is actually in state 0 with probability $P(0)>P(1)$] and finally assumed to be in state 0 (while it is actually in state 1), cf.\ (\ref{workaverage}). As illustrated in Fig. \ref{figure2},  Eq. (\ref{je}) is in good agreement with the numerical results of unitary dynamics (solid line and filled circles, respectively). When the coupling strength to the heat baths is increased, parametrized by $\Gamma^\downarrow \equiv \Gamma^\downarrow_{probe} +\Gamma^\downarrow_{dark}$, the approximation of unitary dynamics during the drive is not anymore valid and deviations from Eq. (\ref{je}) emerge.

Based on similar arguments as done in the derivation of  Eq. (\ref{je}), one can also find from Eq. (\ref{workaverage}) an expression for the distribution of the measured work (again for weak coupling to the baths)
\begin{figure}[b]
\centerline{\includegraphics[width=.6\columnwidth]{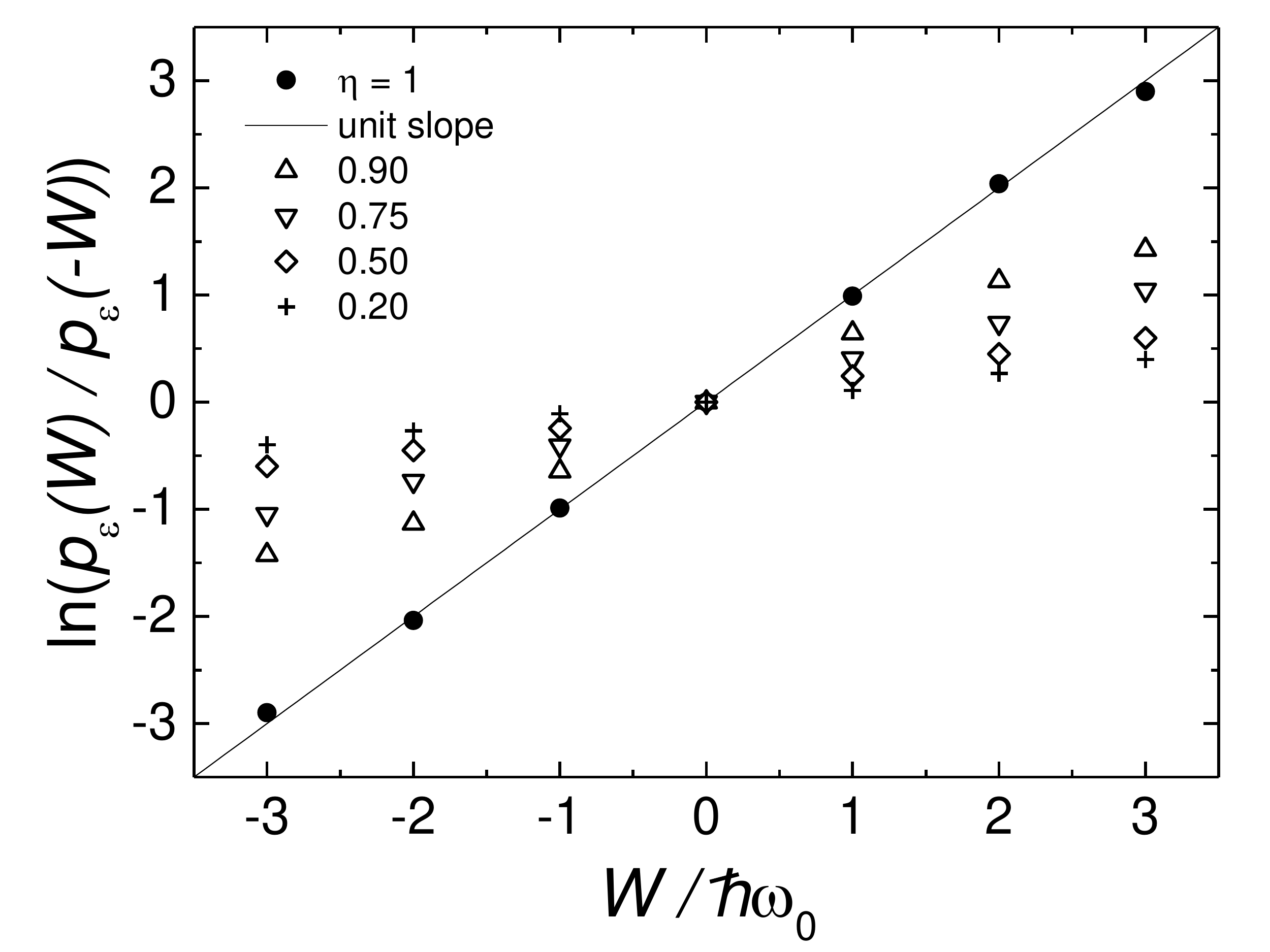}}
\caption{The logarithmic ratio $\ln\left\{{p_\epsilon(\W)}/{p_\epsilon(-W)}\right\}$ in an incomplete measurement. The parameters are $\Gamma^\downarrow = 0.1 \hbar\omega_0$ and $\beta\hbar\omega_0 =1$ and the driving protocol is the same as in Fig. \ref{figure2}. The values of $\eta$ are indicated in the figure.}
\label{figure3}
\end{figure}
\begin{eqnarray}
p_\epsilon(W) &=& \left[ (1-2 \epsilon) P(0) + \epsilon^2\right]\, \delta(W-\hbar\omega_0)+2\epsilon(1-\epsilon) \delta(W)\nonumber\\
&&+\left[ (1-2 \epsilon) P(1) +\epsilon^2\right]\, \delta(W+\hbar\omega_0)\, .
\end{eqnarray}
This allows to write the Crooks relation \cite{crooks99} for the incomplete measurement as
\begin{equation}
\ln\left\{\frac{p_\epsilon(\hbar\omega_0)}{p_\epsilon(-\hbar\omega_0)}\right\}= \ln\left\{\frac{(1-2 \epsilon) P(0) + \epsilon^2}{(1-2 \epsilon) P(1) +\epsilon^2}\right\}\, ,
\end{equation}
illustrating that the Crooks relation is also affected by incomplete measurement. In the limit $\epsilon\to 0$, the standard result of the $\pi$-pulse dynamics is obtained. Figure \ref{figure3} illustrates the effect of $\eta$ on the logarithmic ratio $\ln\left\{p_\epsilon(\W)/{p_\epsilon(-W)}\right\}$ for a coupling strength corresponding to $\Gamma^\downarrow=0.1 \hbar\omega_0$. As can be seen from Fig. \ref{figure3}, the incomplete measurement makes the logarithmic ratio $\ln\left\{p_\epsilon(\W)/{p_\epsilon(-W)}\right\}$  clearly non-linear. In the limit $\epsilon\to 0$, $\ln\left\{p_\epsilon(\W)/{p_\epsilon(-W)}\right\}$ becomes almost linear with the slope given by the Crooks equality, as expected.

\section{Mutual information for the two reservoir setup}\label{mutualinfo}

As we have shown above, the presence of an unaccessible heat bath spoils the measurement of the TLS via an observable reservoir.
This imperfect measurement is thus due to an incomplete information about the probed object which is, in fact, the compound consisting of the TLS and the dark heat bath. Here, we further quantify this lack of information by analyzing the mutual information between the results obtained from the probe reservoir about the state of the TLS  and the actual state of the TLS.

\subsection{Single photon detection}\label{incompletesingle}
The state dependent mutual information between a quantum observable $X$ and its measured value $Y$ is defined as \cite{cover91,sagawa12}
\begin{equation}\label{mutual}
I(x,y) =\ln\left[\frac{P(y|x)}{P(y)}\right]=\ln\left[\frac{P(x|y)}{P(x)}\right]\, ,
\end{equation}
where $P(y|x)$ is the conditional probability to detect $y$ when the true state of the quantum system is $x$ and $P(x)$ is the probability to find the system in $x$. It is related to the joint probability via
\begin{equation}\label{probjoint}
P(x,y)=P(y|x)\, P(x)= P(x|y)\, P(y)
\end{equation}
and to the probability of the detector to measure $y$
\begin{equation}\label{probdetect}
P_D(y)=\sum_{x} P(x,y)\, P(x)\, .
\end{equation}

Now, let us consider the situation discussed above of a TLS coupled to a probe and a dark heat bath. By way of example, we first focus on a single measurement and then turn to the two measurement protocol applied for the work measurement. In the former case one has for the detector probability
\begin{equation}
P_D(0)= (1-\epsilon) \, P(0) + \epsilon\, P(1)\ , \ P_D(1)=\epsilon\, P(0)+ (1-\epsilon)\, P(1)\, ,
\end{equation}
and further, one derives from (\ref{probdetect}) that
\begin{equation}
P(k,0)= \epsilon\, P(k)\ , \ P(k,k)= (1-\epsilon)\, P(k)\ \ , \ k=0,1
\end{equation}
and from (\ref{probjoint}) that
\begin{equation}
P(1|0)=P(0|1)= \epsilon\ \ , \ \ P(k|k)= (1-\epsilon)\ \ , \ k=0,1\, .
\end{equation}
The state dependent mutual information (\ref{mutual}) is then given by
\begin{eqnarray}
I(1,0) &=&\ln\left[\frac{\epsilon}{P(0)}\right]\ ,\ I(0,1)=\ln\left[\frac{\epsilon}{P(1)}\right]\nonumber\\
I(0,0) &=&\ln\left[\frac{1-\epsilon}{P(0)}\right]\ , \ I(1,1) =\ln\left[\frac{1-\epsilon}{P(1)}\right]
\end{eqnarray}
and its average, the mutual information $\langle I\rangle =\sum_{x,y} P(x,y) I(x,y)$, reads
\begin{eqnarray}
\langle I\rangle_\epsilon &=& (1-\epsilon)\left\{ P(0)\, \ln\left[\frac{1-\epsilon}{P(0)}\right]+P(1) \, \ln\left[\frac{1-\epsilon}{P(1)}\right]\right\}\nonumber\\
&&+\epsilon \left\{ P(0)\, \ln\left[\frac{\epsilon}{P(1)}\right]+P(1) \, \ln\left[\frac{\epsilon}{P(0)}\right]\right\}\, .
\end{eqnarray}
For a perfect measurement $\epsilon=0$, the mutual information thus reduces to the entropy of the TLS, i.e.\ $\langle I\rangle_{\epsilon=0}=-P(0) \ln[P(0)]-P(1)\ln[P(1)]$, while in the opposite limit of a completely spoiled detection, $\epsilon \rightarrow 1$, one has  $\langle I\rangle_1= -P(1) \ln[P(0)]-P(0)\ln[P(1)]$. Another limiting case is the domain of high temperatures, where $P(0)\approx P(1)\approx 1/2$ so that
\begin{equation}
\langle I\rangle_\epsilon \approx \ln (2) + (1-\epsilon) \ln(1-\epsilon) + \epsilon \ln(\epsilon)\, .
\end{equation}
Toward zero temperature $P(0)\approx 1$ and $P(1)\ll 1$ we arrive at
\begin{equation}
\langle I\rangle_\epsilon\approx  (1-\epsilon) \ln(1-\epsilon)+\epsilon \ln(\epsilon) -\epsilon \ln[P(1)]
\end{equation}
which for any finite $\epsilon$ is dominated by the rare events when the TLS resides in the '1' state while the prediction assumes that it is in the '0' state.

\subsection{Two photon detection: work measurement}\label{incompletework}

We now turn to the work measurement which, as described above, requires the detection of two photons, the last before the drive and the first after the drive. In both cases, the detector operates not ideally due to the presence of the dark reservoir.

The detector probability $P_D(k_i, k_f)$ [cf.~(\ref{workdetect})] is related to the joint probability that initially the true state of the TLS is $k$ while $k_i$ is detected and that it is finally $k'$ while $k_f$ is detected
\begin{equation}\label{work1b}
P_D(k_f, k_i)= \sum_{k, k'=0,1} P(k_f, k_i; k', k)\, 
\end{equation}
which leads to a generalized conditional probability
\begin{equation}\label{work2}
P(k_f, k_i; k', k)=P(k_f, k_i|k',k)\, P(k)\, p(k\to k')\, .
\end{equation}
According to (\ref{workdetect}) this implies
\begin{equation}
P(k_f, k_i|k', k)= [(1-\epsilon) \delta_{k, k_i}+\epsilon (1-\delta_{k, k_i})]\, [(1-\epsilon) \delta_{k', k_f}+\epsilon (1-\delta_{k', k_f})] \, .
\end{equation}

\begin{figure}[t]
\centerline{\includegraphics[width=.6\columnwidth]{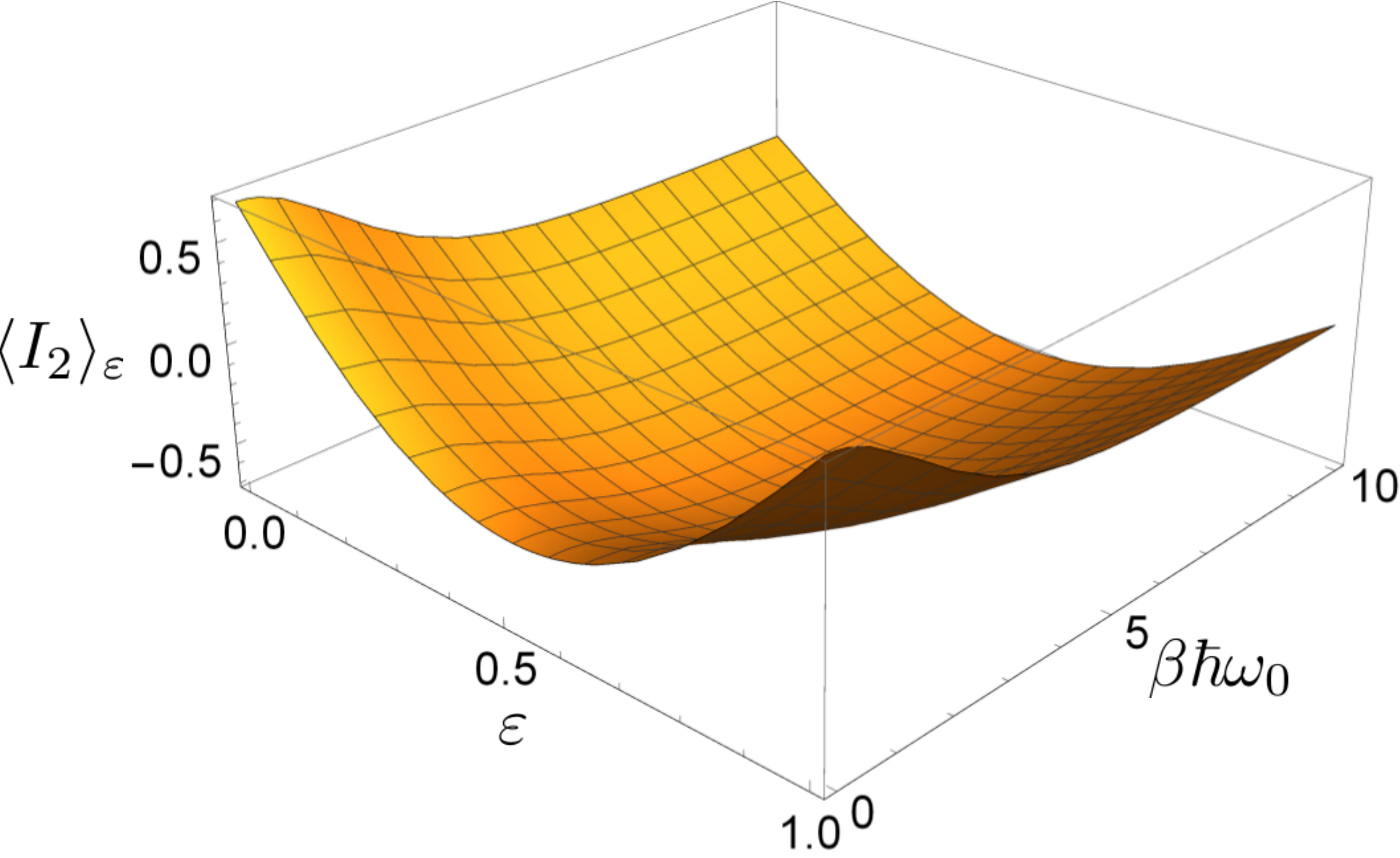}}
\caption{Generalized mutual information $\langle I_2\rangle_\epsilon$ according to Eq. (\ref{mutualinfo2}) vs. the measurement error probability $\epsilon$ and the inverse temperature $\beta \hbar \omega_0$ for a TLS with level spacing $\hbar\omega_0$.}
\label{figure4}
\end{figure}

We now define in generalization of (\ref{mutual}) a state dependent mutual information for the two point measurement
\begin{equation}\label{mutual2}
I_2(k_f,k';k_i,k)={\rm ln}\frac{P(k_f, k_i|k', k)}{P(k_i)p(k_i\to k_f)}\, .
\end{equation}
Its mean $\langle I_2\rangle$ for the swap process is then given by
\begin{eqnarray}
\langle I_2\rangle_\epsilon &=& (1-\epsilon)^2 {\rm ln}(1-\epsilon)^2 +\epsilon^2 {\rm ln}\epsilon^2\nonumber\\
&&-[(1-\epsilon)^2+\epsilon^2]\left\{P(0){\rm ln}[P(0)]+P(1) {\rm ln}[P(1)]\right\}\, .\label{mutualinfo2}
\end{eqnarray}
Note that this expression is symmetric around $\epsilon=1/2$, as illustrated in Fig. \ref{figure4}. It reduces to the entropy of the TLS for an ideal measurement $\epsilon=0$ as well as for a completely spoiled detection $\epsilon\to 1$.

This allows us to formulate together with (\ref{work1b}) and (\ref{work2}) a generalized fluctuation relation which accounts for the incomplete information about the TLS appearing in (\ref{je}) as a deviation from the Jarzynski relation. Namely,
\begin{eqnarray}
\langle {\rm e}^{-\beta W}{\rm e}^{-I_2}\rangle &=& \sum_{k_f, k_i; k',k=0,1} {\rm e}^{-\beta\hbar\omega_0(k_f-k_i)}
\, {\rm e}^{-I_2(k_f,k';k_i,k)} \, P(k_f, k_i; k', k)\nonumber\\
&=& \sum_{k_f, k_i=0,1}{\rm e}^{-\beta\hbar\omega_0 (k_f-k_i)} \,P(k_i)p(k_i\to k_f) \sum_{k, k'=0,1} P(k)\, p(k\to k')\nonumber\\
&=&\sum_{k_f, k_i=0,1} {\rm e}^{-\beta\hbar\omega_0 (k_f-k_i)}\, P(k_i)p(k_i\to k_f)=1\,
\end{eqnarray}
where we used $p(k\to k')=p(k'\to k)$ (micro-reversibility). This verifies that the generalized state dependent mutual information as defined in (\ref{mutual2}) compensates for the incomplete measurement such that the average of the combined expression again obeys a fluctuation relation.


\section{Fast electron-thermometry for calorimetric single-photon detection}

Now we turn our attention to the experimental status of the calorimetric measurement. 
For investigating heat transport and its statistics in small
quantum systems, it is essential to have a highly sensitive
detector with wide bandwidth.
The lack of fast thermometers and calorimeters
in mesoscopic structures has limited
the study of thermodynamics in them.
The variety of phenomena to be explored in thermal physics
would greatly expand with the development of
such devices.
Allowing the detection of temporal evolution of temperatures
under non-equilibrium conditions, fast thermometry would enable
the observation of variations of effective
temperature in small structures as well as the measurements of heat
capacities and energy relaxation rates.
A radio-frequency (RF) electron thermometer
with promise for ultra-low energy calorimetry was presented in Ref.~\cite{gasparinetti14}.
Here we report the latest progress in optimizing the device.

\subsection{Measurement technique and characterization}

\begin{figure}[b]
\centering
\includegraphics[width = 11cm]{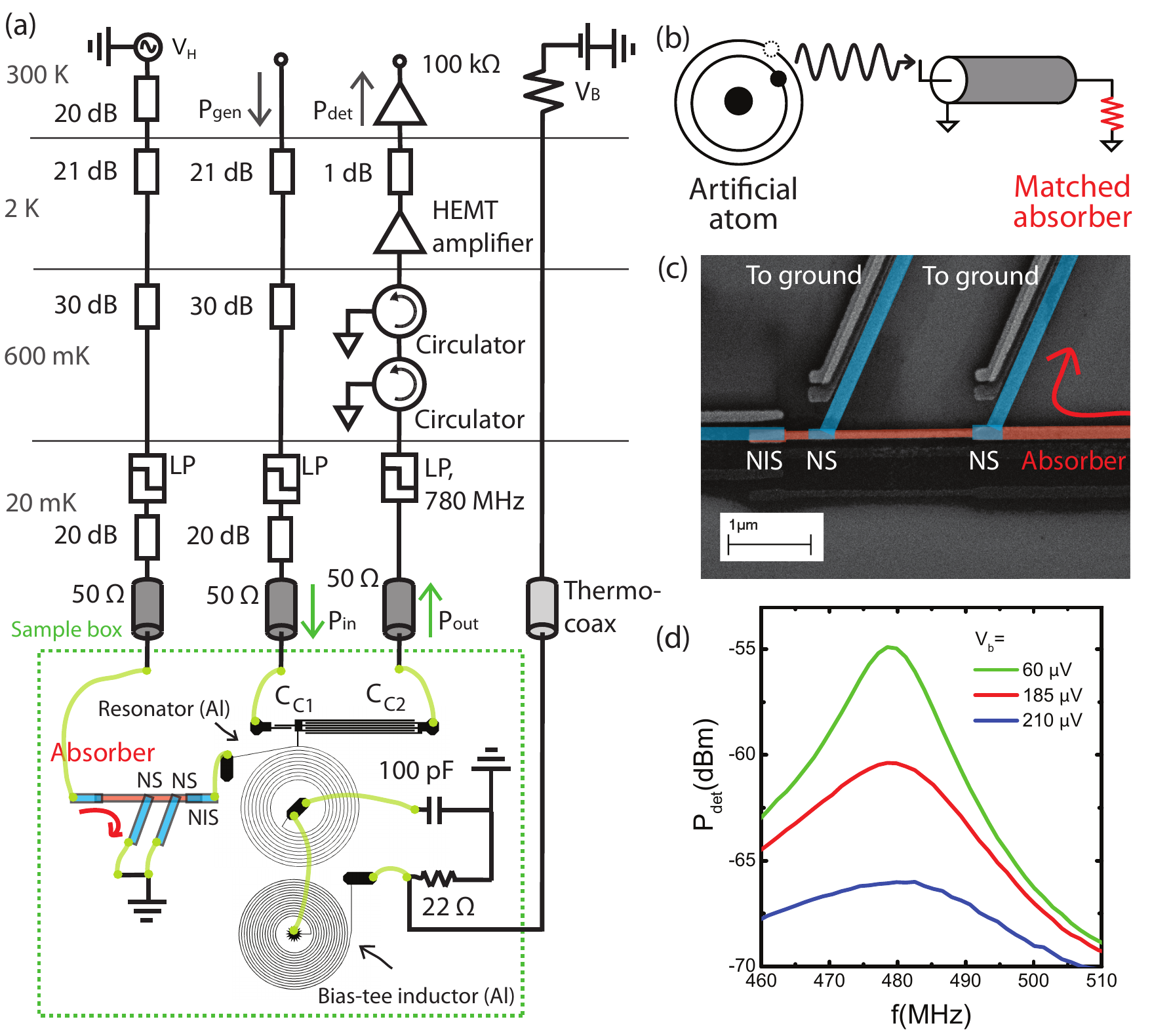}
\caption{(a) Schematics of the measurement setup. The green lines in the sample box illustrate Al bond wires. (b) The scheme of how an artificial atom, e.g. a superconducting qubit, would be connected to an absorber to be measured calorimetrically. (c) SEM image of the sample with false-color highlight on the Cu island (orange) and the Al leads (blue). (d) Resonator lineshape measured at three different values of bias voltage $\Vb$ across the probe junction.}
\label{circuit}
\end{figure}

Our thermometer is currently operating around \SI{100}{mK} electron
temperature ($\Te$).
It is mounted in a
sealed copper box at the cold finger of a dilution
refrigerator.
The measurement circuit is schematically illustrated
in Fig.~\ref{circuit}(a).
The measurement requires a combination of
rf and dc voltages to be applied to the sample.
For high frequency filtering, the dc voltage is
applied through resistive thermocoax cables, while the rf
signal is sent through high frequency coaxial lines.
When examining local temperature of
a small structure, the size of the thermometer
becomes an important figure of merit. 
For calorimetry, it is beneficial to limit the size of the thermometer
for decreasing the heat capacity
of the absorber, and thus the energy resolution
of the detector.
We are using a normal metal-insulator-superconductor (NIS)-tunnel
junction as the temperature-sensitive element
\cite{rowell76, nahum93, giazotto06}. A false-color micrograph
of the sample is shown in Fig.~\ref{circuit}(b).
The overlap area of the NIS-junction is \SI{0.03}{\micro m^2} and the total
volume of the normal metal island
is $\mathcal{V}= 4.5 \cdot 10^{-21}$ m$^3$.
The Cu island is connected to the ground of the sample box
by two Al leads via direct NS-contacts with the normal metal.
Also an rf-line is connected to the island with a direct Al contact
for applying short voltage pulses to heat the sample.
The sample is fabricated on top of an oxidised silicon
substrate by using electron beam lithography,
three-angle metal evaporation and liftoff.

In the standard dc
configuration, the bandwidth of the NIS-thermometer is limited
to the kHz range by the  $\sim 1$ nF capacitance of the measurement cables
and the high differential resistance of the junction. 
For enabling fast readout above MHz range, we have embedded
the junction in an LC resonator, as illustrated in
Fig.~\ref{circuit}(a).
The resonator is made of Al
and is fabricated with a similar method as the sample,
with zero-angle metal evaporation.
The measurement is done in a transmission mode, in which the
NIS-junction is connected to the input and output ports
via capacitors $\Cin$ and $\Cout$.
The transmittance at resonance, $\stos~=~\Pout/\Pin$,
is affected by the temperature dependent conductance
$G$ of the junction as\be
|s_{21}| = 2 \kappa \frac{G_0}{G+G_0} \ , \label{eq:s21}
\ee
with $\kappa = \Cin\Cout/(\Cin^2+\Cout^2)$ and $G_0 = 4 \pi^2 (\Cin^2+\Cout^2) Z_0 f_0^2$. Here $Z_0=\SI{50}{\ohm}$ is the transmission line impedance and $f_0~=~479$ MHz is the resonance frequency. The values of the coupling capacitors are
$\Cin~=~\SI{0.02}{pF}$ and $\Cout~=~\SI{0.4}{pF}$, and
$G_0~=~\SI{67}{\micro S}$. The readout is most sensitive
for differential resistances of the order of $G_0^{-1}~=~\SI{15}{k\Omega}$.

The electron temperature $\Te$ can be estimated from
the transmission measurement by using the
calibrated parameters $\kappa$ and $G_0$.
The conductance of the NIS-junction
can be written as
\be
G =  \frac1{\RT \kB \Te} \int dE N_S(E) f(E-e\Vb) \left[ 1 -
f(E-e\Vb)\right] \ , \label{eq:G}
\ee
where $N_S(E)=\left|\Re{\rm e} \left(E/\sqrt{E^2-\Delta^2}\right)\right|$ is the normalized Bardeen-Cooper-Schrieffer superconducting density of states, $f(E)=\left[1+\exp(E/\kB\Te)\right]^{-1}$
is the Fermi-Dirac function at temperature $T_e$, $k_B$ is the Boltzmann constant, $e$ the electron charge,
$\RT$ is the tunnelling resistance of the junction, and $\Delta$ is the superconducting gap.
For our sample, the parameters are $\RT~=~\SI{9.9}{k\Omega}$
and $\Delta~=~\SI{0.21}{meV}$.
DC bias voltage $\Vb$ is applied to the NIS junction
through a spiral inductor made with the same process
as the resonator.
Due to the bias dependent cooling of the Cu island by the NIS-junction \cite{giazotto06},
$\Te$ varies in the range 85-100~$\mathrm{mK}$ between different bias values within the gap region at the base temperature
of the cryostat, $\Tb~=~\SI{20}{mK}$.  
The bandwidth of the detector, evaluated at the high
differential resistance
region of the NIS-junction, is \SI{10}{MHz} and even higher
at smaller differential resistances.
In Fig.~\ref{circuit}(c), the detected power $\Pdet$
is shown as a function of frequency $f$ at three different values
of $\Vb$.

\subsection{Sensitivity and time resolved measurements}

\begin{figure}[b]
\centering
\includegraphics[width=10cm]{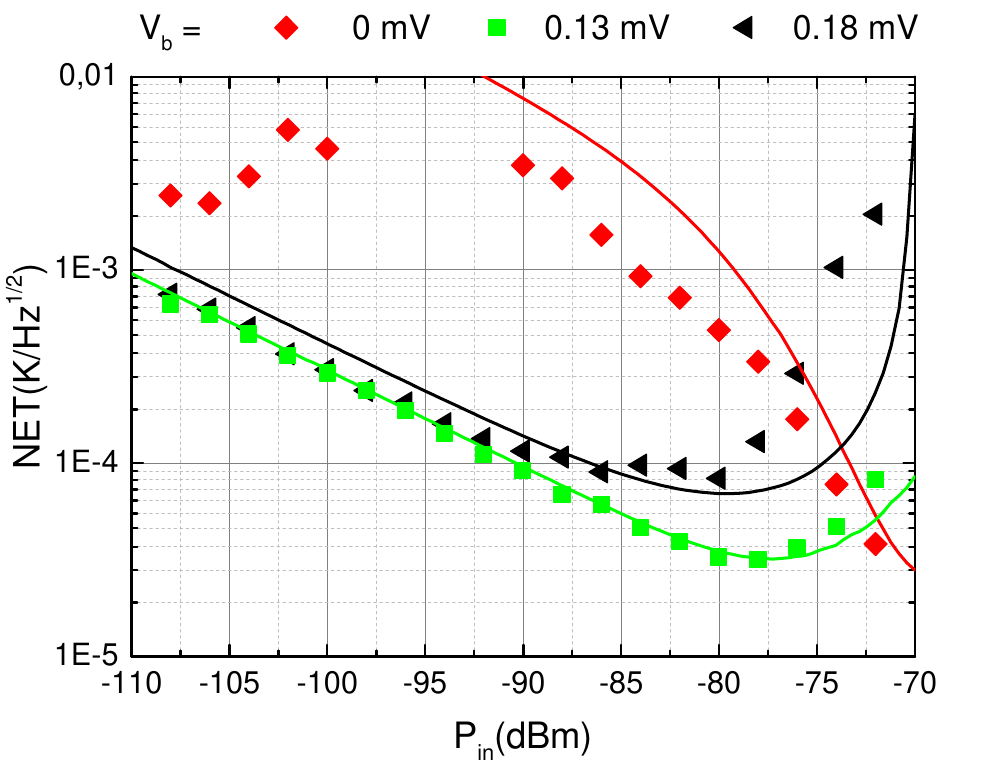}
\caption{Noise equivalent temperature, NET, of the thermometer at selected values of the voltage bias $\Vb$. This measurement was performed at $T_{\rm bath}=\SI{230}{mK}$. The points are measured data. The solid lines are obtained by a numerical simulation using the harmonic balance method to determine the response to a sinusoidal excitation of the resonator terminated by the junction. The decrease of NET observed at the small input power for $\Vb~=~0$ comes from a small supercurrent flowing through the NIS junction, due to the proximity effect induced by the close-by direct NS contact [see Fig. \ref{circuit}(c)].}
\label{net}
\end{figure}

We have evaluated the noise equivalent temperature (NET) of the
thermometer as $ \sqrt{S_{P_{\rm det}}}\mathcal{R}^{-1}$, where $\mathcal{R}~=~\delta P_{\mathrm{det}}/\delta T $ is the responsivity
of the thermometer and $S_{P_{\rm det}}$ is the measured noise spectral density of the detected power $\Pdet$. We obtain $\mathcal{R}$
by measuring $\Pdet$ over a range of bath temperatures $\Tb$ and
evaluating $\mathcal{R}~=~\delta P_{\mathrm{det}}/\delta \Tb $.
In Fig.~\ref{net}, the NET of the detector
is shown as a function of $P_{in}$ at three
selected values of $\Vb$.
The sensitivity of the thermometer is peaked
in a narrow voltage range slightly below
the superconducting gap.
Since the instantaneous voltage across the junction
is a combination of the dc bias and the rf drive,
good sensitivity can be obtained at a variety
of values of $\Vb~<~\Delta /e=\SI{0.21}{\milli V}$, assuming one chooses
a suitable $P_{in}$. This behaviour is confirmed
by the measured data (Fig.~\ref{net}, points) and a numerical simulation (Fig.~\ref{net}, solid lines).
The noise in the measurement was essentially white
and determined by the amplifier.
By characterizing the measurement setup with a system noise parameter $\Tsys$, we can write
$S_{P_{\rm det}} \approx 4 G_{\rm d} k_B T_\mathrm{sys} P_\mathrm{det}$. Here $G_{\rm d} = P_\mathrm{det}/P_\mathrm{out}$ is the total gain of the amplification chain. 
The gain was estimated to be $G_{\rm d}~=~\SI{63}{dB}$. Combined with a noise
measurement, this gives an estimate $\Tsys~=~\SI{13}{K}$.
The best sensitivity we measure
is ${\rm NET}=$ \SI{31}{\micro K/\sqrt{Hz}}. The NET of the thermometer is improved by factor three compared to our previous setup \cite{gasparinetti14}. This is obtained by using a new sample box with improved matching to the 50 $\Omega$ transmission line, a superconducting on-chip resonator and a tunnel junction with lower $R_{\rm T}$. 
The theoretical limit for a fully optimized rf-NIS-thermometer
is ${\rm NET_{\mathrm{opt}}} = \sqrt{2.72 \mathrm{e}^2 T_\mathrm{sys} \RT/k_\mathrm{B}}$,
which in our current setup would be ${\rm NET_{\mathrm{opt}}}~=~\SI{26}{\micro K/\sqrt{Hz}}$.

\begin{figure}[b]
\centering
\includegraphics[width=12cm]{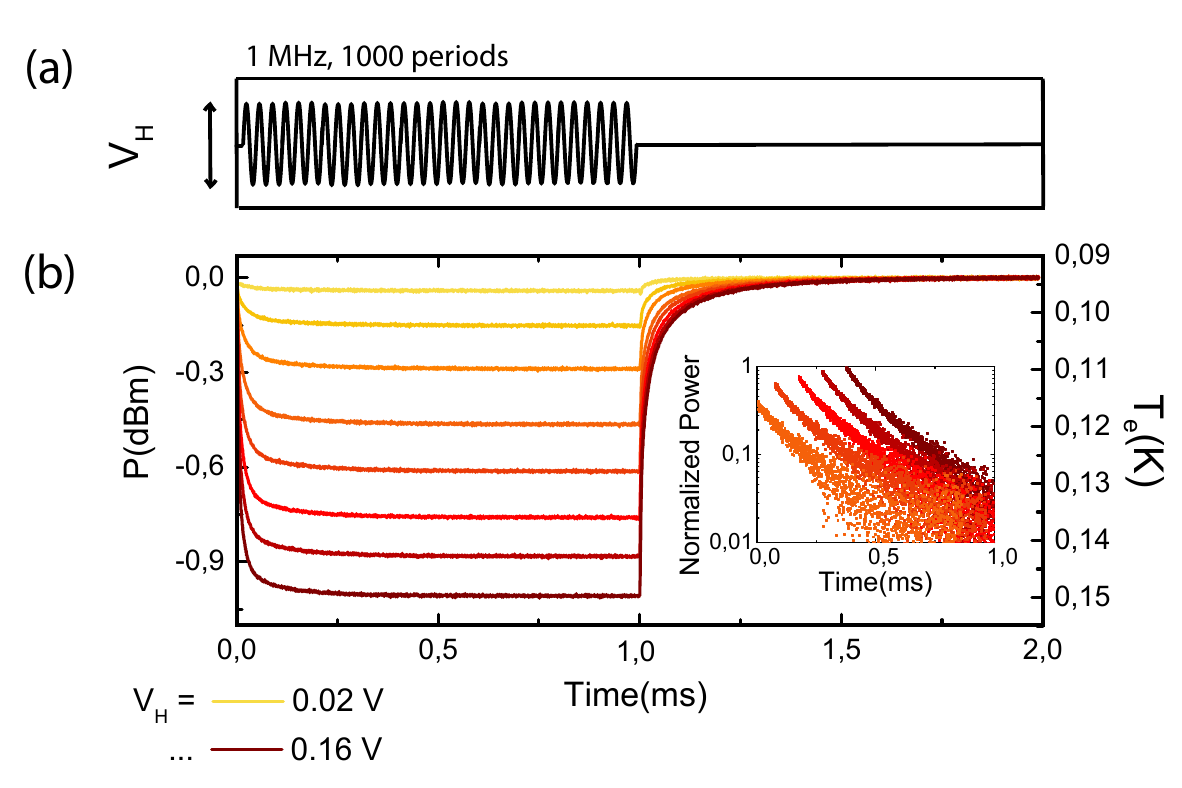}
\caption{Temperature relaxation measurements. (a) The heating pulse consists of a number of periods of \SI{1}{MHz} sinusoidal drive.
(b) Thermometer response to voltage pulses, averaged over 10 000 heating cycles. The power level (P) is shifted
by using the measured power at the steady state as a reference. The different colors correspond to measurements
with different values of $\V_{\mathrm{H}}$ ranging from \SI{0.02}{V} to \SI{0.16}{V}. The relaxation
traces, \SI{100}{\micro s} after the pulse, are shown in the
inset  for a few selected $V_{\mathrm{H}}$. The tails are plotted subtracting the baseline on a logarithmic scale.}
\label{traces}
\end{figure}

For calorimetry, the most important figure of merit is the
energy resolution of the detector,
$\delta E~=~\mathcal{C}\delta T~=~\mathrm{NET}\mathcal{C}\tau ^{-1/2}$.
The smaller the heat capacity $\mathcal{C}$ of the absorber, the larger the temperature change
produced by a single photon absorption event is. Hence, decreasing
the size of the island and choosing a suitable absorber material
is essential in improving the device.
The thermal relaxation time
of the normal metal electrons to the thermal bath is also an important parameter of the calorimeter.
In this work, the temperature relaxation of the Cu island is measured
after heating the normal metal
with a current pulse.
The heating is applied through an rf line, which is
connected to the sample
as illustrated in Figs.~\ref{circuit}(a,c). (In the actual photon counting experiment, this Joule heating will be replaced by pulses from an artificial atom connected via a transmission line to the detector, as shown in Fig.~\ref{circuit}(b).) Short pulses of sinusoidal drive at \SI{1}{MHz} frequency
are used to heat the normal electrons. The form of the pulse is illustrated
in Fig.~\ref{traces}(a), and the response of the thermometer to the
heating is shown in Fig.~\ref{traces}(b). At $\Tb~=~\SI{20}{mK}$,
the thermal relaxation time ($\tau$) is $\sim \SI{100}{\micro s}$
over a wide range of biases at $\Vb~<~\Delta /e$.
The value of $\mathcal C$ can be estimated with the standard expression for a Fermi electron gas, $\mathcal C=\gamma \mathcal V T_{e,0}$, where $\gamma = 71$ JK$^{-2}$m$^{-3}$ \cite{roberts78}.
Together with the measured NET and $\tau$, this
gives for the current setup an estimate
$\delta E /h~=~\SI{4}{THz}$. 
In order to achieve a sufficiently small $\delta E$
for detecting \SI{1}{K} photons of frequency \SI{20}{GHz},
our number needs to be improved.
Since the noise in the measurement is amplifier limited,
the NET of the detector can be improved by
choosing an amplifier with a lower noise temperature
at the first stage. One such choice is a
Josephson parametric amplifier \cite{castellanos-beltran07}.
When the noise in the measurement is limited by thermal fluctuations
on the island as ${\mathrm{NET_{therm}}} = \sqrt{4\kB\Te^2/\Gth}$,
the energy resolution of the detector is
given by $\delta E = \sqrt{4\kB\mathcal{V}\gamma ^2/(5\Sigma\tau)}$.
For the detection of \SI{1}{K}
photons with \SI{100}{\micro s} relaxation time, this would
require to limit the size of a Cu island
to below $7\cdot 10^{-22}$ m$^3$. This can be achieved
with modern fabrication methods, but the proximity
of the superconductor might become the limiting factor
when decreasing the size of the island.
The strength of the thermal coupling between the
electrons and phonons decreases significantly
at lower temperatures resulting in a longer relaxation
time. Also the heat capacity decreases with temperature.
Hence, $\Te$ plays an important role in optimizing the device.
Temperatures of the order of $\SI{10}{mK}$ - almost an order of magnitude smaller than in our current setup - have recently
been measured with an NIS-thermometer \cite{feshchenko14}.

\clearpage
\section{Conclusion}\label{conclusion}
We have discussed some aspects of measuring work and heat in a dissipative two-level quantum system. In the theoretical section, we have analyzed a configuration, where only part of the system and its environment are accessible to the measurement. Including the  counting efficiency of the measurement in the discussion, we have produced modified fluctuation relations. The counting efficiency of the measurement can be associated to the mutual information. We have incorporated the mutual information and recovered general fluctuation relations in a spirit proposed by Sagawa and Ueda for systems with information feedback \cite{sagawa10}. Although our analysis is limited to the situation where the measured and the dark reservoirs have the same temperature, it can easily be generalized to the case of different reservoir temperatures. The results can also be generalised  for $n$-level systems with $n>2$ as long as the instantaneous state of the system after a transition can unequivocally be determined from the energy of the exchanged photon. A notable exception is the harmonic oscillator for which the knowledge of the last transition is not enough to determine the state of system, instead the whole history of transitions is needed due to the equally spaced energy spectrum. 

In the experiment, we have demonstrated an electronic thermometer, operating below \SI{100}{mK}, with \SI{31}{\micro K/\sqrt{Hz}} noise-equivalent temperature and 10 MHz bandwidth.
The device can be integrated into superconducting circuits
with promise for ultralow-energy calorimetry for
mesoscopic structures.
Provided the necessary optimization steps are
taken, our detector will enable calorimetric measurements of single
microwave photons in superconducting quantum circuits.

\section*{Acknowledgments}

We acknowledge discussions with Tapio Ala-Nissila, Michele Campisi and Frank
Hekking. J.A. acknowledges the kind hospitality of the Low Temperature Laboratory at Aalto University. This work has been
supported in part by the European Union Seventh Framework Programme INFERNOS
(FP7/2007-2013) under grant agreement no. 308850, and by Academy of Finland (projects
250280, 251748 and 272218). S.S. acknowledges financial support from the V\"ais\"al\"a
Foundation.

\bibliographystyle{iopart-num.bst}
\section*{References}

\end{document}